\documentclass[aip,jcp,reprint,showkeys,superscriptaddress]{revtex4-2}
\usepackage{graphicx,dcolumn,bm,xcolor,multirow,amscd,amsmath,amssymb,amsfonts,physics,longtable,wrapfig,bbold,xspace,microtype,siunitx}
\usepackage[version=4]{mhchem}

\usepackage[utf8]{inputenc}
\usepackage[T1]{fontenc}

\usepackage{hyperref}
\hypersetup{
    colorlinks,
    linkcolor={red!50!black},
    citecolor={red!70!black},
    urlcolor={red!80!black}
}

\usepackage{listings}
\definecolor{codegreen}{rgb}{0.58,0.4,0.2}
\definecolor{codegray}{rgb}{0.5,0.5,0.5}
\definecolor{codepurple}{rgb}{0.25,0.35,0.55}
\definecolor{codeblue}{rgb}{0.30,0.60,0.8}
\definecolor{backcolour}{rgb}{0.98,0.98,0.98}
\definecolor{mygray}{rgb}{0.5,0.5,0.5}

\definecolor{sqred}{rgb}{0.85,0.1,0.1}
\definecolor{sqgreen}{rgb}{0.25,0.65,0.15}
\definecolor{sqorange}{rgb}{0.90,0.50,0.15}
\definecolor{sqblue}{rgb}{0.10,0.3,0.60}

\lstdefinestyle{mystyle}{
    backgroundcolor=\color{backcolour},
    commentstyle=\color{codegreen},
    keywordstyle=\color{codeblue},
    numberstyle=\tiny\color{codegray},
    stringstyle=\color{codepurple},
    basicstyle=\ttfamily\footnotesize,
    breakatwhitespace=false,
    breaklines=true,
    captionpos=b,
    keepspaces=true,
    numbers=left,
    numbersep=5pt,
    numberstyle=\ttfamily\tiny\color{mygray},
    showspaces=false,
    showstringspaces=false,
    showtabs=false,
    tabsize=2
  }

  \newcolumntype{d}{D{.}{.}{-1}}

\newcommand{\SupMat}{\textcolor{blue}{supporting information}\xspace}
\newcommand{\tabc}[1]{\multicolumn{1}{c}{#1}}
\newcommand{\mr}{\multirow}
\newcommand{\e}{\epsilon}
\newcommand{\Om}{\Omega}
\newcommand{\IP}{I^N}
\newcommand{\EA}{A^N}
\newcommand{\EgFun}{E_\text{g}^\text{fun}}

\usepackage[most]{tcolorbox}
\usepackage{algorithmicx}
\usepackage{algorithm}
\usepackage{algpseudocode}

\newcommand{\LCPQ}{Laboratoire de Chimie et Physique Quantiques (UMR 5626), Universit\'e de Toulouse, CNRS, UPS, France}

\begin{document}	

\title{The $GW$ Approximation: A Quantum Chemistry Perspective}

\author{Antoine \surname{Marie}}
	\email{amarie@irsamc.ups-tlse.fr}
	\affiliation{\LCPQ}
\author{Abdallah \surname{Ammar}}
	\email{aammar@irsamc.ups-tlse.fr}
	\affiliation{\LCPQ}
\author{Pierre-Fran\c{c}ois \surname{Loos}}
	\email{loos@irsamc.ups-tlse.fr}
	\affiliation{\LCPQ}

\begin{abstract}
We provide an in-depth examination of the $GW$ approximation of Green's function many-body perturbation theory by detailing both its theoretical and practical aspects in the realm of quantum chemistry. First, the quasiparticle context is introduced before delving into the derivation of Hedin's equations. From these, we explain how to derive the well-known $GW$ approximation of the self-energy. In a second time, we meticulously explain each step involved in a $GW$ calculation and what type of physical quantities can be computed. To illustrate its versatility, we consider two contrasting systems: the water molecule, a weakly correlated system, and the carbon dimer, a strongly correlated system. Each stage of the process is thoroughly detailed and explained alongside numerical results and illustrative plots. We hope that the contribution will facilitate the dissemination and democratization of Green's function-based formalisms within the computational and theoretical quantum chemistry community.
\bigskip
\begin{center}
	\boxed{\includegraphics[width=0.25\linewidth]{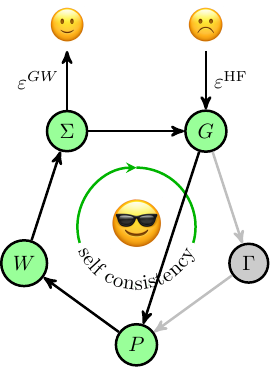}}
\end{center}
\end{abstract}


\maketitle

\section{Quasiparticles}
\label{sec:intro}

The concept of quasiparticles stands as the cornerstone of many-body perturbation theory, serving as a vital tool for characterizing the intricate behaviors of particles within a complex quantum many-body system. \cite{MattuckBook}
Quasiparticles allow, for example, to elucidate the collective behavior of the underlying particles while staying in a single-particle picture.
This is achieved by ``dressing'' the particles of interest with the complex many-body effects to create fictitious particles.
Despite being treated within a single-particle framework, these quasiparticles encapsulate correlation through their dressing.

To illustrate this, consider the example of the removal or addition of an electron.
In a single-particle framework, like Hartree-Fock (HF) theory, the electrons are perceived as independent entities.
According to Koopmans' theorem, the energy needed to extract one electron from the system is merely the negative of its one-particle energy, denoted as $-\e$.
However, this depiction assumes that the act of withdrawing an electron has no impact on the surrounding electrons, which, of course, is far from accurate.

\begin{figure}
  \includegraphics[width=\linewidth]{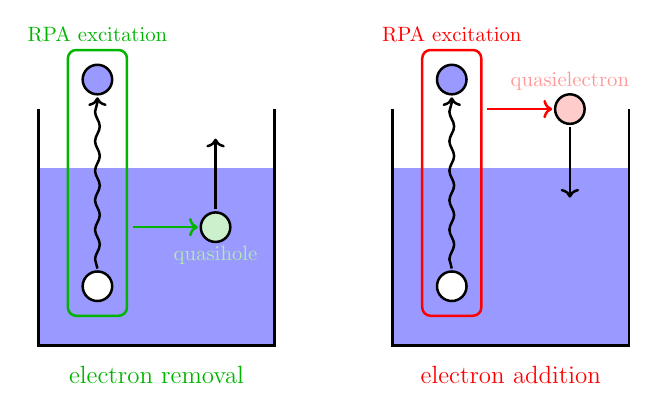}
  \caption{Schematic representation of an electron removal (left) and electron addition (right) process within the $GW$ approximation: the bare hole (white) and the bare electron (blue) are ``dressed'' by the RPA neutral excitations and then become a quasihole (green) and a quasielectron (red).}
  \label{fig:dressing}
\end{figure}

A more realistic perspective portrays the removal of an electron as likely to perturb its neighboring electrons, giving rise to minor disruptions.
These disturbances manifest as neutral excitations, maintaining the total number of particles, yet elevating some electrons to higher-energy one-particle states due to the removal of the targeted electron.
In the quasiparticle framework, these neutral excitations are dressed on the electron transforming it into a ``quasielectron'' with distinct properties compared to its bare counterpart (as illustrated in Fig.~\ref{fig:dressing}).
For example, the interaction between two quasielectrons is characterized by a screened Coulomb interaction because the dressing effectively ``shields'' or ``screens'' each particle.

Consequently, in a single-particle treatment, the energy required to extract this quasielectron becomes $-\e - \Sigma$, where $\Sigma$ represents the so-called self-energy of the quasiparticle.
Thus, even when quasiparticles are considered independently from the rest of the system, the removal energy takes into account many-body effects between real particles through the contribution of the self-energy.
This concept similarly applies to the electron addition process.

In essence, quasiparticles enable a more accurate portrayal of how particles in a many-body quantum system interact and influence one another, shedding light on their collective behavior and dynamic responses to perturbations.

The $GW$ approximation of Green's function many-body perturbation theory, which is discussed in detail in the following, aims to provide a detailed description of the electronic structure and spectral properties of materials and molecules by utilizing the one-body Green's function. It makes extensive use of the quasiparticle concept via the construction of a dynamically-screened version of the Coulomb interaction. \cite{Hedin_1965} $GW$ has been particularly successful in condensed matter physics and has emerged as a highly valuable tool for computing excited-state properties such as bandgaps and band-edge energy levels. We refer the interested reader to Refs.~\onlinecite{Aryasetiawan_1998,Onida_2002,Reining_2017,Golze_2019} for an in-depth discussion of $GW$ (and beyond). The book of Martin, Reining, and Ceperley \cite{MartinBook} is a particularly valuable resource, especially for those newly venturing into this field (see also Refs.~\onlinecite{CsanakBook,FetterBook}). 

Within this context, we propose a more ``quantum chemical'' perspective on the $GW$ approximation. First, we provide a detailed derivation of the $GW$ equations, as outlined in Sec.~\ref{sec:theory}. Then, we present concrete numerical examples involving typical molecular systems exhibiting both weak and strong correlations (see Sec.~\ref{sec:practice}). The content featured in these sections aims to facilitate the implementation of $GW$ within existing quantum chemistry software, bridging the gap between theory and practical applications.

\section{$GW$ in Theory}
\label{sec:theory}

The primary objective of the $GW$ method is to approximate the exact one-body Green's function
\begin{equation}
  \label{eq:g1}
  G(11') = (-\ii)\mel*{\Psi_0^N}{\Hat{T}[\hpsi(1)\hpsid(1')]}{\Psi_0^N}
\end{equation}
where $\Psi_0^N$ is the exact $N$-electron ground-state wave function, $\Hat{T}$ is the time-ordering operator, and $1 = (t_1,\bx_1) = (t_1,\sigma_1,\br_1)$ is a time-spin-space composite index.
$\hpsi(1)$ and $\hpsid(1')$ are second-quantized annihilation and creation field operators in the Heisenberg picture, respectively, that are related to their counterparts in the Schr\"odinger picture as follows
\begin{equation}
	\label{eq:heisenberg}
        \hpsi (1) = e^{\ii\Hat{H}t_1} \Hat{\psi}        (\vb{x_1}) e^{-\ii\Hat{H}t_1}
\end{equation}
Because the electronic Hamiltonian $\Hat{H}$ is time-independent, it is easy to show that $G(11')$ depends only on the time difference $t_1 - t_{1'}$.

The one-body Green's function gives access to the charged excitation energies of the system through its poles in the complex plane as can be readily seen by its Lehmann representation
\begin{equation}
\begin{split}
  G(\bx_1\bx_{1'};\omega) 
  & = \sum_S\frac{\mathcal{I}_S(\bx_1) \mathcal{I}_S^*(\bx_{1'})}{\omega - (E_0^N - E_S^{N-1}) - \ii\eta} 
  \\
  & + \sum_S\frac{\mathcal{A}_S(\bx_1) \mathcal{A}_S^*(\bx_{1'})}{\omega - (E_{S}^{N+1} - E_{0}^N) + \ii\eta} 
\end{split}
\end{equation}
where $\eta$ is a positive infinitesimal, $E_0^N$ is the ground-state energy of the $N$-electron system, while $E_S^{N+1}$ and $E_S^{N-1}$
are the $S$th excited-state energies of the $(N-1)$- and $(N+1)$-electron systems, respectively.
The numerators
\begin{subequations}
\begin{align}
  \mathcal{I}_S(\bx) &= \mel*{\Psi_S^{N-1}}{\hpsi(\bx)}{\Psi_0^{N}}
  \\
  \mathcal{A}_S(\bx) &= \mel*{\Psi_0^N}{\hpsi(\bx)}{\Psi_S^{N+1}}
\end{align}
\end{subequations}
are known as the Lehmann amplitudes or Dyson orbitals.

The success of the $GW$ approximation (and of any other approximations based on Green's functions) hinges on the ability to compute $G$ without explicit reference to the many-body wave function.
This remarkable property is made possible through a closed set of equations known as Hedin's equations. \cite{Hedin_1965}
In the following, we give the outline of their derivation, while a more comprehensive derivation is provided as \SupMat.
Note that, here, we rely on four-point quantities to derive Hedin's equations. \cite{Starke_2012,Maggio_2017b,Orlando_2023a,Orlando_2023b}
We refer the interested reader to Refs.~\onlinecite{Romaniello_2012,Golze_2019,MartinBook} for details about the usual two-point version.

The initial step of this derivation consists of writing down the equation of motion of the one-body Green's function
\begin{multline}
  \label{eq:eom_G1}
  \int \dd 3 \qty[\ii \delta(13)\pdv{t_3} - h(13) ] \, G(31') \\
  + \ii \int \dd (232') v(1 2;3 2') G_2(3 2'^+;1' 2^{++}) = \delta(11')
\end{multline} 
This equation establishes a connection between $G$ and the two-body Green's function, defined as
\begin{equation}
  \label{eq:g2}
  G_2(12;1'2') = (-\ii)^2 \mel*{\Psi_0^N}{\Hat{T}[\hpsi(1)\hpsi(2)\hpsid(2')\hpsid(1')]}{\Psi_0^N}
\end{equation}
Here, $h(12)$ is the one-body part of the Hamiltonian, $\delta(12) = \delta(t_1-t_2)\delta(\bx_1-\bx_2)$ is the Dirac delta function, and $1^\pm = (t_1\pm\eta,\bx_1)$. Additionally, the four-point Coulomb interaction is given by
\begin{equation}
  \label{eq:4point_coulomb}
  v(12;34) = \delta(\bx_1-\bx_{3})\frac{\delta(t_1-t_2)}{\abs{\br_1-\br_2}}\delta(\bx_2-\bx_{4})
\end{equation}
The equation of motion associated with $G_2$ would link the two- and three-body Green's functions.
However, this path does not lead to a closed set of equations for $G$, which is our primary objective.

To proceed, we reframe the equation of motion presented in Eq.~\eqref{eq:eom_G1} as a Dyson equation
\begin{equation}
  G(11') = G_0(11') + \int \dd(23) G_0(12) \Sigma(23) G(31')  
\end{equation}
by introducing the self-energy
\begin{multline}
  \Sigma(11') =	\\
  -\ii \int \dd (232'3') v(1 2;3' 2') G_2(3' 2'^{+};3 2^{++}) G^{-1}(31')
\end{multline}
and the non-interacting one-body Green's function
\begin{equation}
  \int\dd 3\qty[\ii \delta(13)\pdv{t_3} - h(13) ] G_0(31') = \delta(11')
\end{equation}
Then, the next step is to express the self-energy in terms of $G$ and the crucial element for this task is the Martin-Schwinger relation \cite{Martin_1959}
\begin{equation}
  \label{eq:eh_schwinger}
  \eval{\fdv{G(11';[U])}{U(2'2)}}_{U=0} = -G_2(12;1'2') + G(11')G(22')
\end{equation}
which express $G_2$ in terms of the derivative of the one-body Green's function with respect to a fictitious external potential $U$.
The equilibrium Green's function is retrieved by taking the limit $U \to 0$, i.e., $G(11';[U=0]) = G(11')$. In the following, to lighten the notations, we omit the functional dependence in $U$ and the corresponding limit.
The second term in the right-hand side of Eq.~\eqref{eq:eh_schwinger} leads to the Hartree (H) part of the self-energy
\begin{equation}
  \Sigma_{\text{H}}(11') = -\ii\delta(11') \int\dd (22') v(12;1'2') G(2'2^{+})
\end{equation}
and the remaining term encapsulates all the exchange-correlation (xc) effects, which reads, after some manipulation, 
\begin{equation}
  \label{eq:xc_selfenergy}
    \Sigma_{\text{xc}}(11') = \ii \int\dd (22'33') G(33') W(12';32) \tilde{\Gamma}(3'2;1'2')
\end{equation}

In Eq.~\eqref{eq:xc_selfenergy}, the dynamically-screened Coulomb interaction $W$ is also defined through a Dyson equation
\begin{multline}
  W(12;1'2') = v(1 2^{-};1' 2') \\
  - \ii \int \dd(343'4') \, W(14;1'4') \, \tilde{L}(3'4';3^+4) \, v(2 3;2' 3')
\end{multline}
which kernel is the ``irreducible'' polarizability
\begin{equation}
\label{eq:LTilde}
  \tilde{L}(12;1'2') = \int\dd(33') G(13)G(3'1')\tilde{\Gamma}(32;3'2')
\end{equation}
The missing ingredient to close Hedin's equations is the four-point ``irreducible'' vertex function 
\begin{multline}
\label{eq:gw_irred_vertex}
  \tilde{\Gamma}(12;1'2') = \delta(12')\delta(1'2) \\
  + \int\dd (33'44') \Xi_{\text{xc}}(13';1'3) G(34)G(4'3') \tilde{\Gamma}(42;4'2')
\end{multline}
where
\begin{equation}
  \Xi_{\text{xc}}(12';1'2) = \fdv{\Sigma_{\text{xc}}(11')}{G(22')}
\end{equation}
is the exchange-correlation kernel.

We are now in a position to derive the $GW$ approximation.
Neglecting the so-called vertex effects by only considering the first term in Eq.~\eqref{eq:gw_irred_vertex}, i.e., $\tilde{\Gamma}(12;1'2')  \approx \delta(12')\delta(1'2)$, yields the following form of the self-energy
\begin{equation}
  \label{eq:gw_selfenergy}
  \Sigma_{\text{xc}}^{GW}(11') = \ii \int\dd (22') G(2^+2') W(2'1;1'2)
\end{equation}
which gives its name to this particular approximation.
It is worth noting that since $\Tilde{\Gamma}$ also plays a role in the polarizability, as seen in Eq.~\eqref{eq:LTilde}, one has the flexibility to choose whether to apply the same approximation or to opt for an alternative. This choice results in different forms of the $GW$ self-energy. \cite{DelSol_1994,Shirley_1996,Schindlmayr_1998,Morris_2007,Shishkin_2007b,Romaniello_2009a,Romaniello_2012,Gruneis_2014,Hung_2017,Maggio_2017b,Cunningham_2018,Vlcek_2019,Lewis_2019a,Pavlyukh_2020,Wang_2021,Bruneval_2021,Mejuto-Zaera_2022,Wang_2022,Forster_2022b} However, the most common and natural approach is to use the same approximation of $\Tilde{\Gamma}$ for both the self-energy and the polarizability. 
In this case, the irreducible polarizability reads
\begin{equation}
  \label{eq:gw_irred_pol}
  \tilde{L}(1'2';12) = G(1'2)G(2'1)
\end{equation}
and the correponding reducible polarizability 
\begin{multline}
  \label{eq:gw_red_pol}
    L(12;1'2') = \tilde{L}(12;1'2') 
    \\
    - \ii \int\dd (33'44') \tilde{L}(13;1'3') v(3'4';34) L(42;4'^+2')
\end{multline}
is the well-known random-phase approximation (RPA). \cite{Bohm_1951,Pines_1952,Bohm_1953,Nozieres_1958}  
Consequently, $GW$ can be regarded as the first-order approximation of the self-energy with respect to the screened interaction within which the dynamical screening is computed at the RPA level.

Note that here we focus on Hedin's equation, i.e., a closed set of equations in terms of the screened interaction. 
However, an analogous set of equations in terms of the Coulomb interaction can be derived.
This alternative set is also derived in the \SupMat for the sake of completeness.

\section{$GW$ in Practice}
\label{sec:practice}

In this section, we propose to explain the various processes involved in a $GW$ calculation and what type of physical quantities can be computed, and at which step. As examples, we consider one weakly correlated system, the water molecule \ce{H2O}, and one strongly correlated system, the carbon dimer \ce{C2}. Their geometry has been extracted from the \textsc{quest} database. \cite{Loos_2018a,Loos_2019,Veril_2021} By strong correlation, we mean that several electronic states are close in energy to each other (near degeneracy effects). In this case, a multireference treatment might be more appropriate, but one can also employ a single-reference formalism and hope that the post-treatment is accurate enough to compensate for the poor choice of reference configuration. For both closed-shell systems, we consider Dunning's aug-cc-pVTZ basis set. All $GW$ calculations are initiated from HF quantities computed in the restricted formalism. \cite{SzaboBook} All the orbitals are corrected within our scheme and, unless otherwise stated, we set $\eta = 0$. Note that the $GW$ method can be used to correct any mean-field calculations and it is common practice to ``tune'' the starting point using Kohn-Sham orbitals and energies with the ``right'' functional. However, for the sake of simplicity, we shall not consider nor discuss this point in the present review (see, for example, Refs.~\onlinecite{Bruneval_2013,Hung_2017,Gui_2018,Bruneval_2021,Li_2022,McKeon_2022} for illustrative examples).

All the $GW$ calculations reported in the present section have been performed with \textsc{quack}, an open-source software for emerging quantum electronic structure methods, which source code is available at \url{https://github.com/pfloos/QuAcK}.
Implementations of $GW$ methods for localized basis sets are available in several software, such as \textsc{fiesta}, \cite{Blase_2011,Blase_2018} \textsc{bedeft}, \cite{Duchemin_2020,Duchemin_2021} \textsc{molgw}, \cite{Bruneval_2016} \textsc{turbomole}, \cite{vanSetten_2013,Kaplan_2015,Kaplan_2016,Krause_2017} \textsc{adf}, \cite{Forster_2022b,Forster_2022a,Forster_2021,Forster_2020} \textsc{fhi-aims}, \cite{Caruso_2012,Caruso_2013,Caruso_2013a,Caruso_2013b} \textsc{exciton+}, \cite{Patterson_2010,Patterson_2019,Patterson_2020,Hofierka_2022} and \textsc{pyscf}. \cite{Iskakov_2019,Sun_2020,Scott_2023}

Throughout this section, we assume real spinorbitals which is usually the case in molecular calculations, unless a magnetic field is considered. \cite{Holzer_2019} The indices $i$, $j$, $k$, and $l$ are occupied (hole) orbitals; $a$, $b$, $c$, and $d$ are unoccupied (particle) orbitals; $p$, $q$, $r$, and $s$ indicate arbitrary orbitals; $m$ labels single excitations/deexcitations; and $\mu$, $\nu$, $\lambda$, and $\sigma$ denote basis functions.

\subsection{$GW$ Self-Energy}
\label{sec:GW}

In practice, the first step toward obtaining the self-energy is to compute the polarizability.
As alluded to earlier, the irreducible polarizability $\Tilde{L}$, Eq.~\eqref{eq:gw_irred_pol}, is employed to calculate $W$ within the $GW$ approximation via its link with the reducible polarizability $L$, Eq.~\eqref{eq:gw_red_pol}, constructed by utilizing the eigenvalues and eigenvectors of the RPA problem. \cite{SchuckBook,Chen_2015,Ren_2015} The non-Hermitian RPA linear eigenvalue problem is completely defined by the following Casida-like equations expressed in the basis of excitations ($i \to a$) and deexcitations ($a \to i$)
\begin{equation}
\label{eq:RPA}
	\begin{pmatrix} 
    	\bA & \bB
		\\
    	-\bB & -\bA
	\end{pmatrix}
	\cdot
	\begin{pmatrix} 
    	\bX & \bY 
		\\
    	\bY & \bX 
    \end{pmatrix}
  	\\
	=
	\begin{pmatrix} 
    	\bX & \bY
		\\
    	\bY & \bX
    \end{pmatrix}
	\cdot
	\begin{pmatrix} 
    	\bOm & \bO
		\\
    	\bO & -\bOm
    \end{pmatrix}  
\end{equation}
where the diagonal matrix $\bOm$ contains the positive eigenvalues and the normalization conditions are
\begin{subequations}
\begin{align}
	\boldsymbol{X}^{\text{T}} \cdot \boldsymbol{X} - \boldsymbol{Y}^{\text{T}} \cdot \boldsymbol{Y} &= \boldsymbol{1}
	 \\
	\boldsymbol{X}^{\text{T}} \cdot \boldsymbol{Y} - \boldsymbol{Y}^{\text{T}} \cdot \boldsymbol{X} &= \boldsymbol{0}
\end{align}
\end{subequations}
The matrix elements of the (anti)resonant block $\bA$ and the coupling block $\bB$ read 
\begin{subequations}
\begin{align}
	\label{eq:A}
    A_{ia,jb} & = (\e_{a}-\e_{i}) \delta_{ij}\delta_{ab} + \braket{ib}{aj} 
    \\
	\label{eq:B}
    B_{ia,jb} & =  \braket{ij}{ab}
\end{align}
\end{subequations}
with
\begin{equation}
\label{eq:ERI}
    \braket{pq}{rs} = \iint \frac{\MO{p}(\bx_{1})\MO{q}(\bx_{2})\MO{r}(\bx_{1})\MO{s}(\bx_{2})}{\abs{\br_{1}-\br_{2}}} d\bx_{1}d\bx_{2} 
\end{equation}
the usual electron repulsion integrals in the spinorbital basis.
One can estimate the correlation energy at the RPA level via the ``trace'' or ``plasmon'' RPA formula \cite{Sawada_1957a,Rowe_1968,Scuseria_2008,Furche_2008,Jansen_2010,Angyan_2011}
\begin{equation}
  \label{eq:EcRPA}
  \Ec^\RPA = \frac{1}{2} \Tr( \bOm - \bA )
\end{equation}

In practice, in the absence of instabilities, \cite{Seeger_1977a} the linear eigenvalue problem \eqref{eq:RPA} has particle-hole symmetry which means that the eigenvalues are obtained by pairs $\pm \Om_{m}$. Hence, Eq.~\eqref{eq:RPA} can be recast as a Hermitian problem of half its original dimension thanks to the positive definiteness of $\bA-\bB$:
\begin{equation}
\label{eq:small-LR}
	(\bA - \bB)^{1/2} \cdot (\bA + \bB) \cdot (\bA{}{} - \bB{}{})^{1/2} \cdot \bZ{}{} = \bZ{}{} \cdot \bOm^2
\end{equation}
with 
\begin{subequations}
\begin{align}
	\bX + \bY & = (\bA - \bB)^{+1/2} \cdot \bZ \cdot \bOm^{-1/2}
	\\
	\bX - \bY & = (\bA - \bB)^{-1/2} \cdot \bZ  \cdot \bOm^{+1/2}
\end{align}
\end{subequations}
(See the \SupMat for a detailed derivation of these expressions.)
At the RPA level, instabilities are quite unusual. Indeed, in the case of real spin orbitals, we have $\braket{ib}{aj} = \braket{ij}{ab}$, which yields $(\bA-\bB)_{ia,jb} = (\e_{a}-\e_{i}) \delta_{ij}\delta_{ab}$. Therefore, except in the case of unusual orbital occupations, $\bA - \bB$ is positive definite. Alternatively, one can enforce the Tamm-Dancoff approximation by setting $\bB = \bO$ but, in this case, one gets $\Ec^\RPA = 0$, as readily seen in Eq.~\eqref{eq:EcRPA}.

Using these quantities, one can compute the elements of the dynamically screened Coulomb interaction as 
\begin{equation}
\label{eq:W}
	W_{pq,rs}(\omega)
	= \braket{pq}{rs} 
	+ \sum_{m} \qty[ 
	\frac{M_{pr,m} M_{qs,m}}{\omega - \Om_m + \ii\eta}
	-\frac{M_{pr,m} M_{qs,m}}{\omega + \Om_m - \ii\eta}
	]
\end{equation}
where the transition densities (or ``screened'' two-electron integrals) are given by
\begin{equation}
	\label{eq:sERI}
	M_{pq,m} = \sum_{jb} \braket{pj}{qb} \qty(X_{jb,m} + Y_{jb,m})
\end{equation}
Due to the spin structure of the $GW$ equations, only the singlet RPA excitations contribute to the transition densities, which is used in practice to further reduce the size of the RPA problem via spin adaptation. \cite{Angyan_2011,Orlando_2023b}

It is important to note that one needs to compute the entire spectrum of eigenvalues and their corresponding eigenvectors, as the sum over $m$ in Eq.~\eqref{eq:W} cannot be truncated to the first few excitations, as commonly performed in linear response calculations via Davidson's algorithm. \cite{Davidson_1975} (We shall get back to this point later on.) Because the RPA matrices [see Eq.~\eqref{eq:RPA}] are of size $\order*{N^2}$ (where $N$ is the number of orbitals), this step costs $\order*{N^6}$ and is thus the computational bottleneck of a $GW$ calculation. Important techniques, such as the contour deformation approach, have been designed to leverage this. \cite{vanSetten_2013,Neuhauser_2013,Govoni_2015,Liu_2016,Vlcek_2017,Wilhelm_2018,Duchemin_2019,DelBen_2019,Forster_2020,Duchemin_2020,Duchemin_2021,Forster_2021,Panades-Barrueta_2023} (See Ref.~\onlinecite{Golze_2019} for an extensive list of the different techniques.)

\begin{figure*}
	\includegraphics[width=\linewidth]{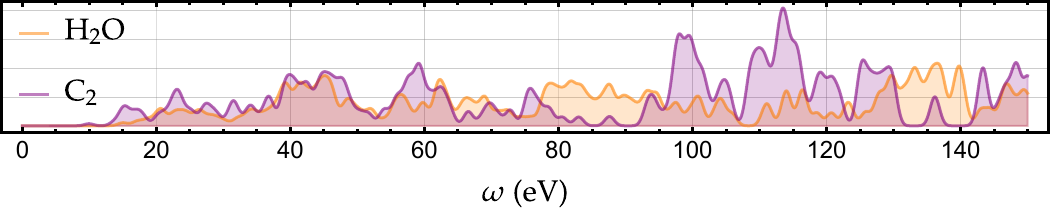}
	\caption{Low-frequency part of the absorption spectrum of \ce{H2O} (orange) and \ce{C2} (purple) computed at the RPA level with the aug-cc-pVTZ basis. See main text for more details.}
	\label{fig:eps}
\end{figure*}

Figure \ref{fig:eps} shows the low-frequency part of the absorption spectrum of \ce{H2O} and \ce{C2} computed at the RPA/aug-cc-pVTZ level.
It has been modeled as a convolution of Gaussian functions centered at each (singlet) neutral excitation frequency $\Om_m$ with intensity proportional to the corresponding oscillator strength of the transition: \footnote{A broadening parameter of \SI{1}{\eV} has been applied.}
\begin{equation}
	f_m = \frac{2}{3} \Om_m \norm{\boldsymbol{\mu}_m}^2
\end{equation}
where the transition dipole moment is
\begin{equation}
	\boldsymbol{\mu}_m = \sum_{ia} \mel{i}{\br}{a} (X_{ia,m}+Y_{ia,m})
\end{equation}
and
\begin{equation}
	\mel{p}{\br}{q} =  \int \MO{p}(\bx) \br \MO{q}(\bx) d\bx
\end{equation}
are one-electron integrals.
Figure \ref{fig:eps} evidences the fine structure of such a quantity and the difficulty of modeling it with the well-known (generalized) plasmon-pole approximation that is widely applied in solid-state calculations, \cite{Larson_2013} but also in molecular systems. \cite{Deslippe_2012,vanSetten_2015}

Finally, performing the convolution of the Green's function and the dynamically screened interaction [see Eq.~\eqref{eq:gw_selfenergy}], the elements of the correlation part of the $GW$ self-energy read
\begin{equation}
	\label{eq:SigGW}
	\Sigma^{\text{c}}_{pq}(\omega) 
	= \sum_{im}\frac{M_{pi,m} M_{qi,m}}{\omega - \e_{i} + \Om_{m} - \ii\eta}
	+\sum_{am}\frac{M_{pa,m} M_{qa,m}}{\omega - \e_{a} - \Om_{m} + \ii\eta}
\end{equation}
Note that, in a $GW$ calculation, it is not mandatory to compute explicitly the elements of $W$ defined in Eq.~\eqref{eq:W}.
Only the transition densities [see Eq.~\eqref{eq:sERI}] are required to compute the self-energy elements in Eq.~\eqref{eq:SigGW}. 

At this stage, we have not specified the nature of the orbitals and their energies in Eqs.~\eqref{eq:A}, \eqref{eq:B}, \eqref{eq:ERI}, and \eqref{eq:SigGW} as this choice depends on the level of self-consistency one is willing to include.

\subsection{Level of Self-Consistency}
\label{sec:sc}

The practical equation to be solved is known as the quasiparticle equation 
\begin{equation}
	\label{eq:quasipart_eq}
	\left[\bF + \bSig^{\text{c}}(\omega=\epsilon_p)\right]\psi_p(\bx) = \epsilon_p \psi_p(\bx)
\end{equation}
where $\epsilon_p$ are the quasiparticle energies and $\psi_p(\bx)$ are the corresponding (Dyson) orbitals. $\bF$ is the usual Fock matrix and $\bSig^{\text{c}}(\omega)$ is a matrix with elements derived from Eq.~\eqref{eq:SigGW}. This equation can be interpreted as the HF equation augmented with an additional term that accounts for correlations arising from the screening effect among electrons. Note also the close link with Kohn-Sham density-functional theory. However, the primary challenge in solving this equation stems from the frequency-dependent nature of the self-energy, rendering it a nonlinear and non-Hermitian matrix equation. As a result, it is common practice to employ approximations to solve the quasiparticle equation.

In the popular one-shot scheme, known as $G_0W_0$, \cite{Strinati_1980,Hybertsen_1985a,Godby_1988,Linden_1988,Northrup_1991,Blase_1994,Rohlfing_1995} one only considers the diagonal part of the self-energy and performs a single iteration of Hedin's equations (see pseudo-code in Fig.~\ref{fig:pseudo}). 
Considering a HF starting point, the quasiparticle energies are thus obtained by solving the non-linear quasiparticle equation for each orbital $p$
\begin{equation}
\label{eq:qp}
	\omega - \e_{p}^\HF = \Sigma^{\text{c}}_{pp}(\omega)
\end{equation}
where $\e_{p}^\HF$ are the HF one-electron energy of the $p$th orbital.

It is also practically convenient to linearize this quasiparticle equation by performing a first-order Taylor expansion of the self-energy around the starting point energy, i.e., $\omega = \e_{p}^\HF$, 
\begin{equation}
	\Sigma^{\text{c}}_{pp}(\omega) 
	\approx \Sigma^{\text{c}}_{pp}(\e_{p}^{\HF}) 
	+ (\omega - \e_{p}^{\HF}) \eval{\pdv{\Sigma^{\text{c}}_{pp}(\omega)}{\omega}}_{\omega = \e_{p}^{\HF}}
\end{equation}
yielding
\begin{equation}
	\e_{p} = \e_{p}^\HF + Z_{p} \Sigma^{\text{c}}_{pp}(\omega = \e_{p}^\HF)
\end{equation}
where the value of the renormalization factor $Z_{p}$ gives the spectral weight of the corresponding quasiparticle solution $\e_{p}$
\begin{equation}
	\label{eq:Z}
	\qty( Z_{p} )^{-1}
	= 1 - \eval{\pdv{\Sigma^{\text{c}}_{pp}(\omega)}{\omega}}_{\omega=\e_{p}^\HF}
\end{equation}
The value of $Z_p$ can easily be shown to be strictly restricted between $0$ and $1$.
This scheme is coined lin$G_0W_0$ in the following and its pseudo-code is given in Fig.~\ref{fig:pseudo}.
When the so-called quasiparticle approximation holds, the weight of the quasiparticle equation is close to unity, while the remaining weight is distributed among the satellite (or shake-up) transitions.

Several levels of self-consistency can be achieved. One common approach is to enforce self-consistency on the quasiparticle energies, a scheme referred to as the ``eigenvalue'' self-consistent $GW$ (ev$GW$) method. \cite{Hybertsen_1986,Shishkin_2007a,Blase_2011,Faber_2011,Rangel_2016} In the ev$GW$ approach, a series of iterations is conducted, during which updates are made solely to the one-electron energies $\e_{p}$ used in defining the RPA matrices [as seen in Eqs.~\eqref{eq:A} and \eqref{eq:B}] and the self-energy [see Eq.~\eqref{eq:SigGW}], while the corresponding orbitals remain unchanged. This represents the simplest and most widely employed self-consistent scheme. The pseudo-code of the ev$GW$ procedure is given in Fig.~\ref{fig:pseudo}.

For a more comprehensive self-consistent treatment, the quasiparticle self-consistent $GW$ (qs$GW$) scheme can be employed, which extends the self-consistency to the orbitals themselves. \cite{Gui_2018,Faleev_2004,vanSchilfgaarde_2006,Kotani_2007,Ke_2011,Kaplan_2016,Forster_2021,Marie_2023} In this approach, both the one-electron energies and the orbitals are iteratively updated until convergence is achieved. The procedure involves diagonalizing an effective Fock matrix that incorporates a correlation potential. This potential is derived from a static Hermitian self-energy, expressed as:
\begin{equation}
  \Tilde{\Sigma}^{\text{c}}_{pq} = \frac{\Sigma^{\text{c}}_{pq}(\e_{p}) + \Sigma^{\text{c}}_{qp}(\e_{q})}{2}
\end{equation}
Recently, we derived, from first-principles, an alternative static Hermitian form for the qs$GW$ self-energy based on the similarity renormalization group (SRG) approach. \cite{Marie_2023} The pseudo-code of the qs$GW$ procedure is given in Fig.~\ref{fig:pseudo}.
Although qs$GW$ performs self-consistency on the orbitals and the quasiparticle energies, it is not strictly independent of the starting point as it is not unusual to obtain different solutions depending on the initial set of orbitals and energies.

One should not confuse qs$GW$ with the fully self-consistent $GW$ (sc$GW$) approach, \citep{Caruso_2012,Caruso_2013a,Caruso_2013b,Koval_2014,Grumet_2018,DiSabatino_2021,Yeh_2022} where one updates the poles and weights of $G$ retaining quasiparticle and satellite energies at each iteration. We shall not address this scheme here.

\begin{figure*}
\begin{tcolorbox}[colback=orange!50!white]
\begin{algorithmic}[1]
\Procedure{$G_0W_0$@HF}{}
\State Perform HF calculation to get orbital coefficients $\bc^\HF$ and energies $\be^\HF$
\State Integral transformation $\braket{\mu\nu}{\lambda\sigma} \xrightarrow[]{\bc^\HF} \braket{pq}{rs}$
\State Construct RPA matrices $\bA$ and $\bB$ from $\be^\HF$ and $\braket{pq}{rs}$, as defined in Eqs.~\eqref{eq:A} and \eqref{eq:B}
\State Compute RPA eigenvalues $\bOm$ and eigenvectors $\bX + \bY$ by diagonalizing the linear eigenvalue problem \eqref{eq:RPA}
\State Construct screened two-electron integrals $M_{pq}^{m}$, as defined in Eq.~\eqref{eq:sERI}
\For {$p=1,\ldots, N$}
    \State Solve the non-linear equation $\omega = \e_{p}^\HF + \Sigma^{\text{c}}_{pp}(\omega)$ using Newton's method starting from $\omega = \e_{p}^\HF$ to get the quasiparticle energy $\e_{p}^{G_0W_0}$
\EndFor
\EndProcedure
\end{algorithmic} 
\end{tcolorbox}
%
\begin{tcolorbox}[colback=yellow!50!white]
\begin{algorithmic}[1]
\Procedure{lin$G_0W_0$@HF}{}
\State Perform HF calculation to get orbital coefficients $\bc^\HF$ and energies $\be^\HF$
\State Integral transformation $\braket{\mu\nu}{\lambda\sigma} \xrightarrow[]{\bc^\HF} \braket{pq}{rs}$
\State Construct RPA matrices $\bA$ and $\bB$ from $\be^\HF$ and $\braket{pq}{rs}$, as defined in Eqs.~\eqref{eq:A} and \eqref{eq:B}
\State Compute RPA eigenvalues $\bOm$ and eigenvectors $\bX + \bY$ by diagonalizing the linear eigenvalue problem \eqref{eq:RPA}
\State Construct screened two-electron integrals $M_{pq}^{m}$, as defined in Eq.~\eqref{eq:sERI}
\For {$p=1,\ldots, N$}
    \State Compute the self-energy element $\Sigma_{pp}(\omega)$ given in Eq.~\eqref{eq:SigGW} at $\omega = \e_{p}^\HF$ 
    \State Compute the renormalization factor $Z_p$ defined in Eq.~\eqref{eq:Z}
    \State Evaluate the quasiparticle energy $\e_{p}^{G_0W_0} = \e_{p}^\HF + Z_{p} \Sigma^{\text{c}}_{pp}(\omega = \e_{p}^\HF)$
\EndFor
\EndProcedure
\end{algorithmic} 
\end{tcolorbox}

\begin{tcolorbox}[colback=red!50!white]
\begin{algorithmic}[1]
\Procedure{ev$GW$@HF}{}
\State Perform HF calculation to get orbital coefficients $\bc^\HF$ and energies $\be^\HF$
\State Integral transformation $\braket{\mu\nu}{\lambda\sigma} \xrightarrow[]{\bc^\HF} \braket{pq}{rs}$
\State Set $\be^{G_{-1}W_{-1}} = \be^\HF$, and $n = -1$
\While{$\Delta>\tau$}
\State $n \gets n+1$
\State Construct RPA matrices $\bA$ and $\bB$ from $\be^{G_{n}W_{n}}$ and $\braket{pq}{rs}$, as defined in Eqs.~\eqref{eq:A} and \eqref{eq:B}
\State Compute RPA eigenvalues $\bOm$ and eigenvectors $\bX + \bY$ by diagonalizing the linear eigenvalue problem \eqref{eq:RPA}
\State Construct screened two-electron integrals $M_{pq}^{m}$, as defined in Eq.~\eqref{eq:sERI}
\For {$p=1,\ldots, N$}
    \State Solve the non-linear equation $\omega = \e_{p}^\HF + \Sigma^{\text{c}}_{pp}(\omega)$ using Newton's method starting from $\omega = \e_{p}^{G_{n}W_{n}}$ to get the quasiparticle energy $\e_{p}^{G_{n+1}W_{n+1}}$
\EndFor
\State $\Delta = \max \abs{ \be^{G_{n+1}W_{n+1}} - \be^{G_{n}W_{n}} }$ 
\EndWhile
\EndProcedure
\end{algorithmic} 
\end{tcolorbox}

\begin{tcolorbox}[colback=purple!50!white]
\begin{algorithmic}[1]
\Procedure{qs$GW$}{}
\State Perform HF calculation to get orbital coefficients $\bc^\HF$ and energies $\be^\HF$
\State Set $\be^{G_{-1}W_{-1}} = \be^\HF$, $\bc^{G_{-1}W_{-1}} = \bc^\HF$, and $n = -1$
\While{$\Delta>\tau$}
\State $n \gets n+1$
\State Integral transformation $\braket{\mu\nu}{\lambda\sigma} \xrightarrow[]{\bc^{G_{n}W_{n}}} \braket{pq}{rs}$
\State Construct RPA matrices $\bA$ and $\bB$ from $\be^{G_{n}W_{n}}$ and $\braket{pq}{rs}$, as defined in Eqs.~\eqref{eq:A} and \eqref{eq:B}
\State Compute RPA eigenvalues $\bOm$ and eigenvectors $\bX + \bY$ by diagonalizing the linear eigenvalue problem \eqref{eq:RPA}
\State Construct screened two-electron integrals $M_{pq}^{m}$, as defined in Eq.~\eqref{eq:sERI}
\State Evaluate the self-energy elements $\Sigma_{pq}^{\text{c}}(\omega)$ given in Eq.~\eqref{eq:SigGW} at $\omega = \e_{p}^{G_{n}W_{n}}$ and form $\Tilde{\bSig}^{\text{c}} = \qty[\bSig^{\text{c}}\T{+(\bSig^{\text{c}})}]/2$
\State Form the Fock matrix $\bF$ in orbital basis from $\bc^{G_{n}W_{n}}$
\State Diagonalize $\bF + \Tilde{\bSig}$ to get $\be^{G_{n+1}W_{n+1}}$ and $\bc^{G_{n+1}W_{n+1}}$
\State $\Delta = \max \abs{ \be^{G_{n+1}W_{n+1}} - \be^{G_{n}W_{n}} }$ 
\State $n \leftarrow n+1$
\EndWhile
\EndProcedure
\end{algorithmic} 
\end{tcolorbox}
\caption{Pseudo-algorithm for each $GW$ scheme: $G_0W_0$@HF, lin$G_0W_0$@HF, ev$GW$@HF, and qs$GW$.}
\label{fig:pseudo}
\end{figure*}

\subsection{Correlation energy}
\label{sec:Ec}

Despite being a one-body quantity, $G$ can be used to compute the energy of the system.
The Galitskii-Migdal (GM) functional, \cite{Galitskii_1958} which reads 
\begin{equation}
  \label{eq:gm_functional}
  E^\text{GM} = -\frac{\ii}{2} \int d\bx_1 \lim_{2\to1^+}\qty[ \ii \pdv{}{t_1} + h(\bx_1) ] G(12)
\end{equation}
gives the total electronic energy of the system.
The correlation energy can thus be extracted from it and is obtained via the convolution of the correlation part of the self-energy and the Green's function as
\begin{equation}
  \Ec^\text{GM} = -\frac{\ii}{2} \int_{-\infty}^{\infty} \frac{\dd\omega}{2\pi} \int \dd \bx_1\bx_3e^{\ii\omega\eta} \Sigma_{\text{c}}(\bx_{1}\bx_{3};\omega) G(\bx_{3}\bx_1;\omega)
\end{equation}
which can be recast as
\begin{equation}
  \label{eq:EcGM}
  \Ec^\text{GM} = - \sum_{iam}\frac{M_{ia,m}^2}{\e_{a} - \e_{i} + \Om_{m}}
\end{equation}
The derivation of the previous equations starting from Eq.~\eqref{eq:gm_functional} is included in the \SupMat.
In the context of $GW$, the Galitskii-Migdal functional, which is known to be non-variational and strongly dependent on the quality of $G$, grossly overestimates the correlation energy in molecular systems. \cite{Stan_2006,Caruso_2012,Caruso_2013,Caruso_2013a,Caruso_2013b,Pokhilko_2022,Pokhilko_2021b,Pokhilko_2021a,Bruneval_2021b}
Meaningful energies are only obtained at full self-consistency \cite{Loos_2018b,DiSabatino_2021} or by adding higher-order terms. \cite{Ren_2013}

The Galitskii-Migdal functional is not the only functional based on $G$ available to compute energies.
There are two well-known alternatives, the Luttinger-Ward \cite{Luttinger_1960} and Klein \cite{Klein_1961} functionals, which become variational if $G$ satisfies a Dyson equation. 
These two functionals are equivalent to the Galitskii-Migdal functional if $G$ is obtained through full self-consistency.
To calculate the correlation energy at the $GW$ level using these functionals, an additional RPA calculation is performed and the correlation energy is evaluated via Eq.~\eqref{eq:EcRPA} using the corresponding $GW$ quantities. \cite{Stan_2006,Dahlen_2006,Dahlen_2005a,Dahlen_2005,Dahlen_2004a,Dahlen_2004,Stan_2009}
(See Ref.~\onlinecite{Caruso_2013b} for a detailed derivation.)

\subsection{Quasiparticle energies}
\label{sec:qp}

\begin{figure}
	\includegraphics[width=\linewidth]{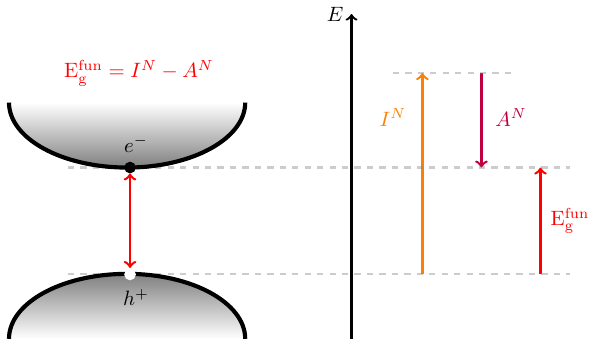}
	\caption{Fundamental gap $\EgFun = \IP - \EA$, where $\IP$ and $\EA$ are the principal ionization potential and the principal electron affinity of the $N$-electron system.}
	\label{fig:Eg}
\end{figure}

Here, we compute the principal ionization potential (IP), $\IP$, principal electron affinity (EA), $\EA$ , and fundamental gap ($\EgFun = \IP - \EA$), as depicted in Fig.~\ref{fig:Eg}.
At the $GW$ level, these quantities are simply given by
\begin{align}
	\IP & \approx -\e_{\HOMO}^{\GW}
	&
	\EA & \approx -\e_{\LUMO}^{\GW}
\end{align}
where HOMO and LUMO stand for the highest-occupied and lowest-unoccupied molecular orbitals, respectively.

\begin{figure*}
	\includegraphics[width=\linewidth]{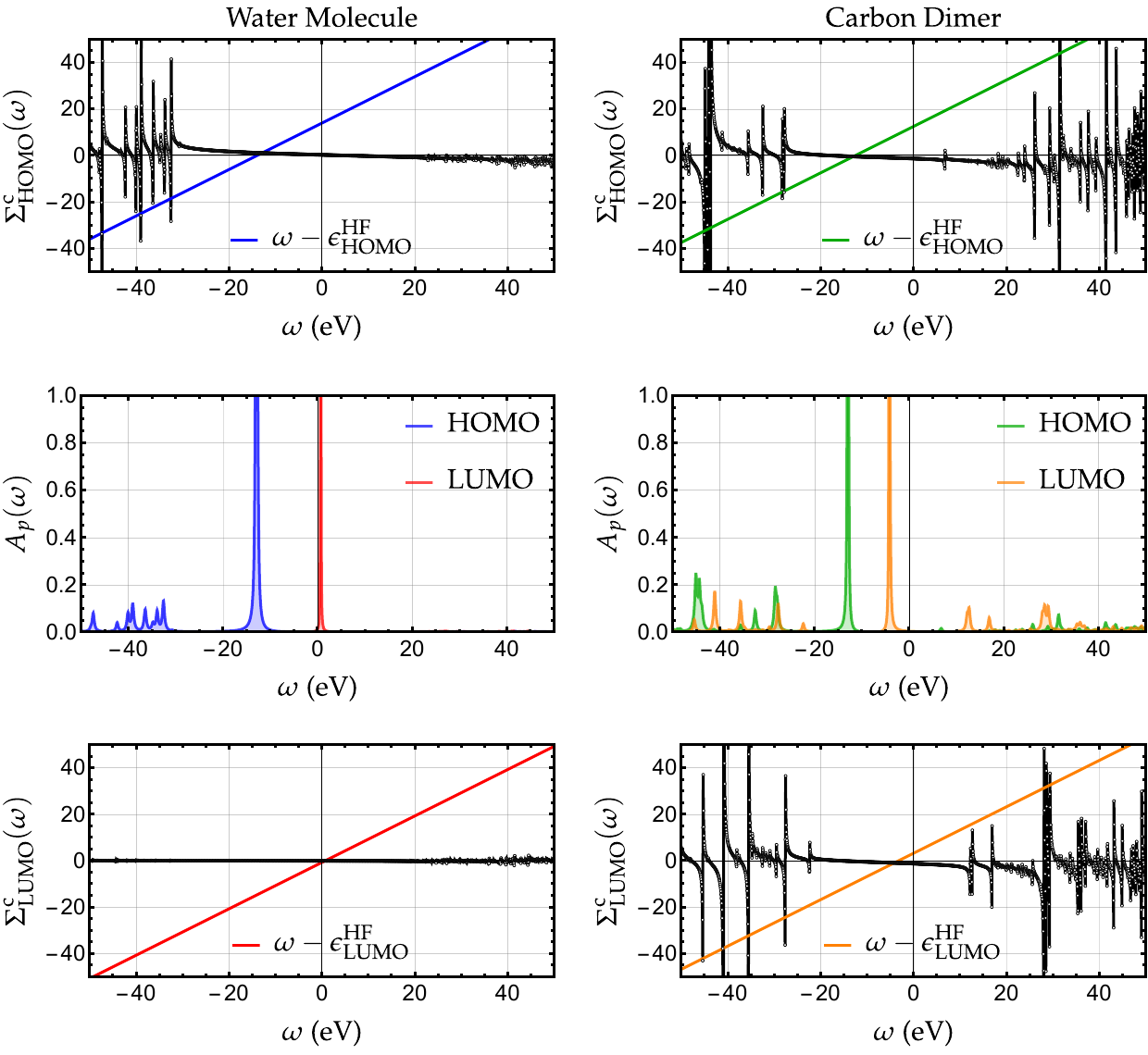}
	\caption{Self-energy (black curves) associated with the HOMO (top) and LUMO (bottom) orbitals of \ce{H2O} (left) and \ce{C2} (right) computed at the $G_0W_0$ level of theory with the aug-cc-pVTZ basis and $\eta = \SI{d-3}{\hartree}$. The solutions of the quasiparticle equation are given by the intersection of the black and colored curves. In the central panels, the spectral function (in \si{\hartree^{-1}}) associated with the HOMO and LUMO orbitals are represented.}
	\label{fig:SigC}
\end{figure*}

The black curves in Fig.~\ref{fig:SigC} correspond to the variation of the $G_0W_0$ self-energy associated with the HOMO and LUMO orbitals of \ce{H2O} and \ce{C2} as a function of $\omega$, while the colored lines correspond to $\omega - \e_{p}^\HF$. Therefore, the solutions of the diagonal quasiparticle equation \eqref{eq:qp} can be found where the black and colored curves intersect. Their respective spectral weight is directly related to the slope of the self-energy at these specific intersections. When the slope is small, the renormalization factor is close to unity [see Eq.~\eqref{eq:Z}] and the solution is categorized as a quasiparticle. In the scenario where the slope is high, the renormalization factor is small and the solution corresponds to a satellite. As illustrated in Fig.~\ref{fig:SigC}, satellite solutions typically originate from poles in the self-energy. In all the cases depicted in Fig.~\ref{fig:SigC}, there is a well-defined quasiparticle solution, clearly separated from the region where the self-energy poles are located. However, it is not uncommon to encounter situations where the weight is almost equally distributed between two solutions. \cite{vanSetten_2015,Maggio_2017,Loos_2018b,Veril_2018,Duchemin_2020,Marie_2024} In such a case, one can refer to a breakdown of the quasiparticle picture.

One can gain a deeper understanding of how spectral weight is distributed among multiple solutions by examining the spectral function linked to each orbital $p$. This spectral function, as evidenced by the following equation, is intricately connected to the imaginary part of the one-body Green's function
\begin{equation}
\label{eq:spectral_function}
\begin{split}
	A_{p}(\omega) 
	& = \frac{1}{\pi} \abs{ \Im G_{pp}(\omega)} 
	\\
	& = \frac{1}{\pi} \frac{\abs{\Im\Sigma_{pp}^{\text{c}}(\omega)}}{ \qty[ \omega - \e_p^\HF - \Re\Sigma_{pp}^{\text{c}}(\omega) ]^2 + \qty[\Im\Sigma_{pp}^{\text{c}}(\omega)]^2}
\end{split}
\end{equation}
Figure \ref{fig:SigC} reports $A_{\HOMO}(\omega)$ and $A_{\LUMO}(\omega)$ for both systems. 
As one can see, it is clear that the quasiparticle solution carries most of the weight in these two specific examples.

\begin{table*}
\caption{Principal IP ($\IP$), principal EA ($\EA$), fundamental gap ($\EgFun = \IP - \EA$), in units of \si{\eV}, and RPA and Galitskii-Migdal correlation energies ($\Ec^\RPA$ and $\Ec^\GM$), in units of \si{\hartree}, for \ce{H2O} and \ce{C2} computed at various levels of theory with the aug-cc-pVTZ basis. The corresponding FCI values and a selection of experimental measurements are reported for the sake of comparison.}
\label{tab:res}
	\begin{ruledtabular}
		\begin{tabular}{llddddrc}
Mol. 	& 	& \tabc{lin$G_0W_0$@HF}  & \tabc{$G_0W_0$@HF} & \tabc{ev$GW$@HF} & \tabc{qs$GW$\fnm[1]}	&	\tabc{FCI\fnm[2]} 	&	\tabc{Exp.\fnm[3]}\\					
		\hline
\ce{H2O}	&	$\IP$		&	12.885	&	12.884	&	12.764	&	12.879	&	$12.679$			&	$12.600$		\\
	 		&	$\EA$		&	-0.685	&	-0.685	&	-0.681	&	-0.662	&	$-0.608$			&					\\
	 		&	$\EgFun$	&	13.570	&	13.569	&	13.446	&	13.541	&	$13.287$			&					\\
	 		&	$\Ec^\RPA$	&	-0.343	&	-0.345	&	-0.348	&	-0.358	&	\mr{2}{*}{$-0.298$}	&					\\
	 		&	$\Ec^\GM$	&	-0.615	&	-0.615	&	-0.631	&	-0.647	&						&					\\
\ce{C2}		&	$\IP$		&	12.928	&	12.928	&	12.953	&	12.561	&	$12.463$			&	$11.4\pm0.3$	\\
			&	$\EA$		&	4.153	&	4.153	&	4.229	&	3.840	&	$3.175$				&	$3.273\pm0.008$	\\
	 		&	$\EgFun$	&	8.775	&	8.775	&	7.914	&	8.722	&	$9.288$				&					\\
	 		&	$\Ec^\RPA$	&	-0.395	&	-0.399	&	-0.401	&	-0.423	&	\mr{2}{*}{$-0.416$}	&					\\
	 		&	$\Ec^\GM$	&	-0.685	&	-0.685	&	-0.699	&	-0.730	&						&					\\
		\end{tabular}
	\end{ruledtabular}
	\fnt[1]{Calculations performed with SRG regularization and a flow parameter $s = 500$, as described in Ref.~\onlinecite{Marie_2023}}
	\fnt[2]{Calculations performed with \textsc{quantum package} (freely available at \url{https://github.com/QuantumPackage/qp2}) using the \textit{``Configuration Interaction using a Perturbative Selection made Iteratively''} (CIPSI) method. \cite{Garniron_2019}}
	\fnt[3]{Values extracted from the \textit{Computational Chemistry Comparison and Benchmark DataBase} (CCCBDB) \cite{CCCBDB} at \url{https://cccbdb.nist.gov}.}
\end{table*}

The $GW$ results for the IP, EA, and fundamental gap obtained at various levels of self-consistency are reported in Table \ref{tab:res}. They are compared with the exact results, computed in the same basis set, through full configuration interaction (FCI) calculations on the cationic, anionic, and neutral species. \cite{Garniron_2019}
Additionally, Table \ref{tab:res} includes the RPA and Galitskii-Migdal correlation energies [see Eqs.~\eqref{eq:EcRPA} and \eqref{eq:EcGM}]. It is clear that the RPA estimates are in much better agreement with the FCI reference values than the Galitskii-Migdal correlation energies which are significantly lower (approximately by a factor of two) and exhibit larger fluctuations with respect to the level of self-consistency, in line with our earlier discussion in Sec.~\ref{sec:Ec}.

As anticipated, the $GW$ estimates for the IP of water are fairly accurate, while the EA of water is found to be negative. Hence, $GW$ is not suited to model such an unstable anion. In contrast, the carbon dimer has a stable anionic state, and although \ce{C2} is a prototypical strongly correlated system characterized by a substantial configuration mixing in the $N$-electron ground state (the weight of the HF reference determinant being only $0.69$ in the ground-state FCI wave function), the IP and EA values obtained at the $GW$ level are satisfactory. \cite{Pavlyuk_2007,Ammar_2024}

A point worth mentioning is that the linearization of the quasiparticle equation \eqref{eq:qp} is usually an outstanding approximation. This is clearly exemplified here with errors below \SI{0.001}{\eV} for \ce{H2O} and \ce{C2}. Actually, if one observes a substantial disparity between the lin$G_0W_0$@HF and lin$G_0W_0$ numbers, it might indicate a potential issue, raising concerns about the validity of the quasiparticle approximation. In such situations, it may be prudent to closely examine the spectral function and check for the presence of additional close-lying solutions with significant weight.


\begin{figure}
	\includegraphics[width=\linewidth]{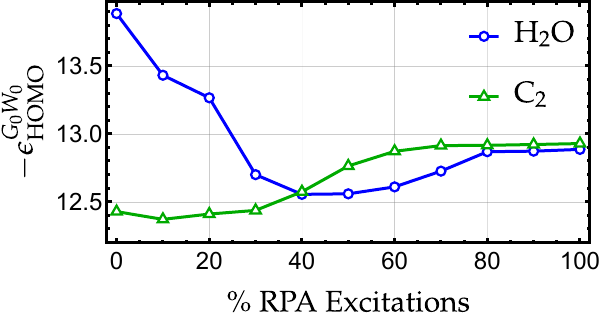}
	\caption{Evolution of the quasiparticle energy associated with the HOMO of \ce{H2O} (blue) and \ce{C2} (green) computed at the lin$G_0W_0$ level of theory with the aug-cc-pVTZ basis with respect to the percentage of RPA excitations taken into account in Eq.~\eqref{eq:W}.}
	\label{fig:W}
\end{figure}

Let us return quickly to the RPA problem. In Sec.~\ref{sec:GW}, we mentioned that one must compute the entire spectrum of eigenvalues and eigenvectors to obtain well-converged quasiparticle energies. This is illustrated in Fig.~\ref{fig:W} where we show the evolution of the quasiparticle energy associated with the HOMO of \ce{H2O} and \ce{C2} with respect to the percentage of RPA excitations taken into account in Eq.~\eqref{eq:W}. From this, it is clear that the convergence is quite erratic and non-monotonic, and it is thus hard to design an approximate model with a limited number of poles for the RPA polarizability.

\begin{figure*}
	\includegraphics[width=0.7\linewidth]{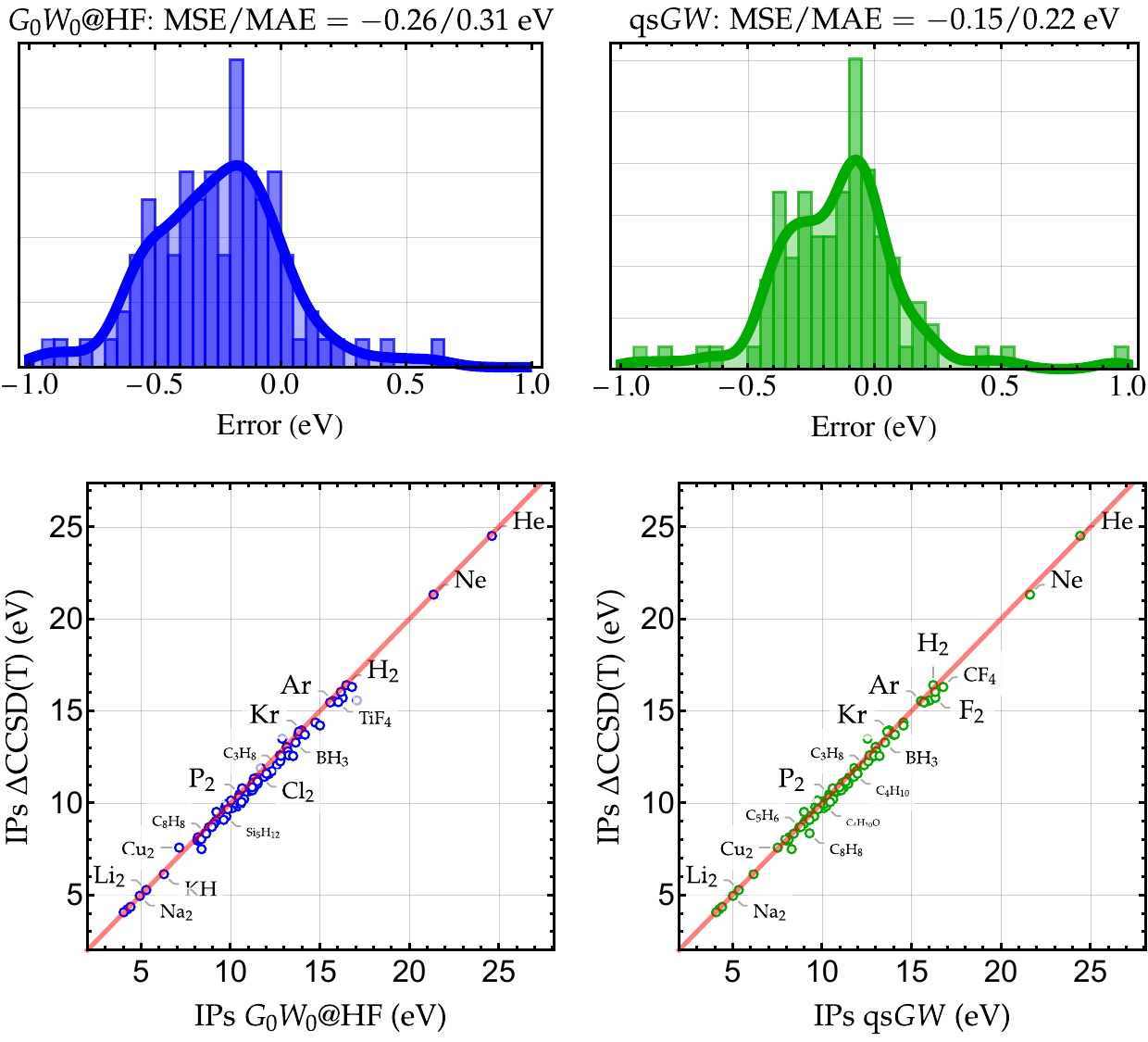}
	\caption{Histograms (top) and scatter plots (bottom) of the errors in the principal IPs of the $GW100$ dataset \cite{vanSetten_2015} computed at the $G_0W_0$@HF and qs$GW$ levels considering the $\Delta$CCSD(T) values of Ref.~\onlinecite{Krause_2017} as reference.
	Data for $G_0W_0$@HF and qs$GW$ have been extracted from Ref.~\onlinecite{Caruso_2016} and are available at \url{https://gw100.wordpress.com}. The mean signed error (MSE) and mean absolute error (MAE) with respect to the reference values are also reported.}
	\label{fig:GW100}
\end{figure*}

Overall, self-consistency has a beneficial effect in finite systems like atoms and molecules, where both partially and fully self-consistent $GW$ methods have exhibited significant promise. \cite{Ke_2011,Blase_2011,Faber_2011,Caruso_2012,Caruso_2013,Caruso_2013a,Caruso_2013b,Koval_2014,Hung_2016,Blase_2018,Jacquemin_2017a,Forster_2022a,Marie_2023} Conversely, the situation is more contentious in solid-state calculations, as the self-consistency and vertex corrections are recognized to offset each other to a certain extent. \cite{MartinBook} The debate surrounding the significance of partial and full self-consistency in the $GW$ method has persisted for a long time. \cite{Stan_2006,Stan_2009,Rostgaard_2010,Caruso_2012,Caruso_2013,Caruso_2013a,Caruso_2013b,Koval_2014,Wilhelm_2018} In certain scenarios, it has been observed that self-consistency can actually degrade spectral properties when compared to the simpler one-shot $G_0W_0$ approach. This phenomenon was notably demonstrated in the context of calculations on the uniform electron gas,  \cite{Holm_1998,Holm_1999,Holm_2000,Garcia-Gonzalez_2001} a fundamental model with relevance to many fields of physics and chemistry. \cite{Loos_2016} Such observations were further confirmed in real extended systems. \cite{Schone_1998,Ku_2002,Kutepov_2016,Kutepov_2017} It is important to acknowledge that other approximations might have contributed to this deterioration, such as the use of pseudo-potentials \cite{deGroot_1995} or finite-basis set effects. \cite{Friedrich_2006} Consequently, these studies have cast doubt on the necessity of employing self-consistent schemes within the $GW$ framework, at least for solid-state calculations.

In Fig.~\ref{fig:GW100}, we report the mean signed error (MSE) and mean absolute error (MAE) associated with the principal IPs of the $GW100$ dataset \cite{vanSetten_2015} computed at the $G_0W_0$@HF and qs$GW$ levels \cite{Caruso_2016} considering the $\Delta$CCSD(T) values as reference. \cite{Krause_2017} The distributions of the errors and the corresponding scatter plots are also provided. Going from $G_0W_0$@HF to qs$GW$ lowers both the MAE from \SI{0.31}{\eV} to \SI{0.22}{\eV} and the MSE from \SI{-0.26}{\eV} to \SI{-0.15}{\eV}. Although the overall underestimation of $GW$ for principal IPs remains, it is significantly reduced via the introduction of self-consistency.

\section{Concluding Remarks}
\label{sec:ccl}

We report an extensive review of the $GW$ approximation within the framework of Green's function many-body perturbation theory, offering, we hope, a comprehensive analysis of both its theoretical foundations and practical applications in the context of quantum chemistry.
As a starter, we introduce the concept of quasiparticles, providing a necessary backdrop for a deep dive into the derivation of Hedin's equations, a crucial starting point for our subsequent discussion on how to derive the well-established $GW$ approximation of the self-energy.

Following this, we meticulously guide the reader through each step involved in a $GW$ calculation, elucidating the panel of physical quantities that can be computed using this approach. To showcase its adaptability and effectiveness, we turn our attention to two distinct systems: the weakly correlated water molecule and the strongly correlated carbon dimer. At each stage of the process, we provide a comprehensive breakdown and offer clear explanations, complemented by numerical results and illustrative plots. The effect of self-consistency on the quasiparticle energies, which is clearly beneficial in the case of molecular systems, has been illustrated on the $GW100$ database.

The ultimate goal of this review is to facilitate the dissemination and democratization of Green's function-based formalisms within the computational and theoretical quantum chemistry community.
Given the numerous successes of many-body perturbation theory across various fields of physics, quantum chemistry can certainly benefit from this formalism. \cite{FetterBook,DickhoffBook}
We refer the interested reader to Refs.~\onlinecite{MartinBook,CsanakBook,FetterBook,Aryasetiawan_1998,Onida_2002,Reining_2017,Golze_2019} for more in-depth discussions around Green's function many-body perturbation theory.

\acknowledgments{
We gratefully acknowledge the discussions over the years with Pina Romaniello, Arjan Berger, Fabien Bruneval, and Xavier Blase. 
This project has received financial support from the European Research Council (ERC) through the European Union's Horizon 2020 
--- Research and Innovation program --- under grant agreement no.~863481. 
Additionally, it was supported by the European Centre of Excellence in Exascale Computing (TREX), and has received funding from 
the European Union's Horizon 2020 --- Research and Innovation program --- under grant agreement no.~952165.}

\bibliography{GWrev}

\end{document}